# ON THE RELEVANCE OF BANDWIDTH EXTENSION FOR SPEAKER VERIFICATION[1]

*Marcos Faúndez-Zanuy, Mattias Nilsson (\*), W. Bastiaan Kleijn (\*)*

Escola Universitària Politècnica de Mataró (UPC) SPAIN

(\*) Department of Speech, Music and Hearing (KTH) SWEDEN
e-mail: faundez@eupmt.es, mattiasn@speech.kth.se, bastiaan@speech.kth.se

## ABSTRACT

In this paper, we consider the effect of a bandwidth extension of narrow-band speech signals (0.3-3.4 kHz) to 0.3-8 kHz on speaker verification. Using covariance matrix based verification systems together with detection error trade-off curves, we compare the performance between systems operating on narrow-band, wide-band (0-8 kHz), and bandwidth-extended speech. The experiments were conducted using different short-time spectral parameterizations derived from microphone and ISDN speech databases. The studied bandwidth-extension algorithm did not introduce artifacts that affected the speaker verification task, and we achieved improvements between 1 and 10 percent (depending on the model order) over the verification system designed for narrow-band speech when mel-frequency cepstral coefficients for the short-time spectral parameterization were used.

## 1. INTRODUCTION

In speaker verification, the objective is to verify a person's claimed identity from his or her voice. Many different techniques exist for performing this task [12]. In general, speaker verification algorithms perform better for wide-band signals than for narrow-band signals. This makes it interesting to analyze the effect of bandwidth-extension algorithms, which recover high- (3.4 –8 kHz) frequency band given a narrow-band speech signal (0.3-3.4 kHz) and, sometimes, additionally the low band (0.1-0.3 kHz). For an ideal verification algorithm, a bandwidth-extended signal should not lead to better or worse verification performance than a narrow-band signal since no independent information about the speaker is introduced. As we will show, conventional verification algorithms are not far from this ideal behavior.

A second motivation for our study is the fact that bandwidth-extension will likely be used commonly in future heterogeneous networks, e.g. [13]. Our results thus add to other known results on the effect of network-based processing on speaker verification, e.g., [1,2,3].

This paper is organized as follows. Section 2 describes the speaker-recognition and the bandwidth-extension algorithms that were used in our studies. Section 3 deals with the database and the simulation procedure. Section 4 presents the results, and section 5 is devoted to the main conclusions.

## 2. SPEAKER VERIFICATION AND BANDWIDTH EXTENSION

### 2.1 Verification using covariance matrices

In this study, we are only interested in the relative performance between systems designed for narrow-band, wide-band, or bandwidth-extended speech. Thus, we have chosen a simple second-order based measure for the verification of a speaker.

In the training phase, we compute for each speaker and each bandwidth scenario empirical covariance matrices based on feature vectors extracted from overlapped short-time segments of the speech signals, i.e., $C_j = \hat{E}[x_n x_n^T]$, where $\hat{E}$ denotes estimate of the mean and $x_n$ represents the features vector for frame *n*. As features representing short-time spectra we use both linear prediction cepstral coefficients (LPCC) and mel-frequency cepstral coefficients (MELCEPST) [5].

In the speaker-verification system, the trained covariance matrices for each speaker are compared to an estimate of the covariance matrix obtained from a test sequence from a speaker. An arithmetic-harmonic sphericity measure is used in order to compare the matrices [6]:

$$d = \log\left(\text{tr}(C_{test} C_j^{-1})\, \text{tr}(C_j C_{test}^{-1})\right) - 2\log(l),$$

where $\text{tr}(\cdot)$ denotes the trace operator, *l* is the dimension of the feature vector, $C_{test}$ and $C_j$ is the covariance estimate from the test speaker and speaker model *j*, respectively. We applied the relation $p=\exp(-0.5d)$ to convert the distance measure *d* into a probability measure *p*. Comparing *p* to a threshold results in a detection decision that is either correct (true speaker accepted), miss (true speaker rejected), or false alarm (impostor accepted). Counting the number of occurances for the error types, miss detection and false alarm, over the test set we obtain the empirical probabilities $P_{miss}$ and $P_{fa}$, respectively.

### 2.2 Bandwidth extension

The bandwidth-extension method used in this work is based on statistical modeling between the narrow- and high-band [4]. The core of this system is a Gaussian mixture model (GMM) that models the true joint probability-density function (pdf) between the narrow- and high-band feature vectors. Both the narrow- and high-band feature vectors consist of a number of parameters describing the spectral envelope of each band. In addition, the narrow-band feature vector has one parameter being a measure on the degree of voicing, and the high-band feature vector includes the log-energy ratio between the two frequency bands.

---

[1] This work has been supported by the CICYT TIC2000-1669-C04-02

A schematic of the bandwidth extension algorithm is depicted in Fig 1. From the narrow-band speech, 15 cepstral coefficients and the degree of voicing are derived. Then, using an asymmetric costfunction, penalizing high-band energy over-estimates, we obtain an estimate of the energy ratio between the high- and narrow-band. Given the estimated energy ratio and the derived narrow-band features, an MMSE estimate of the spectral envelope of the high-band is computed. A modified spectral-folding based excitation is then filtered with the energy-ratio controlled high-band envelope and added to the up-sampled low-pass filtered narrow-band speech signal to form a reconstructed wide-band speech signal.

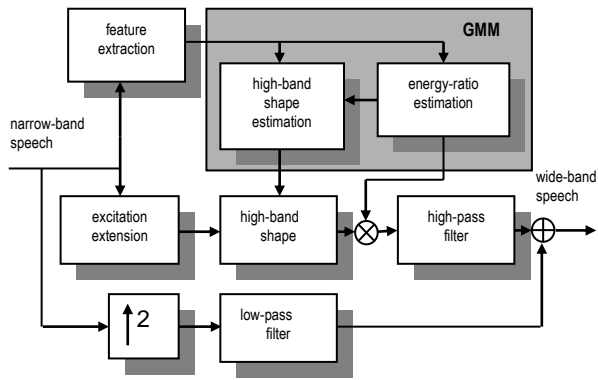

Fig. 1 Bandwidth-extension system architecture.

## 3. SIMULATIONS

### 3.1 Simulation procedure

The speech signals (narrow-band, wide-band, and bandwidth extended) are pre-emphasized by a first order filter whose transfer function is $H(z)=1-0.95z^{-1}$. A 30 ms Hamming window is used (unless another value is stated), and the overlap between adjacent frames is 2/3. One minute of read text is used for training, and five sentences for testing (each sentence is about two seconds long). The experiments were conducted using LPCC and MELCEPST with dimensions ranging from 4 to 27.

The extracted features from the training data are used to estimate a model (covariance matrix) of each speaker for the different bandwidth cases. From each test sentence the sphericity measure is computed for each speaker model and converted to a probability (likelihood) measure and then compared to a threshold as described in the section 2. Lowering the detection threshold, we are decreasing the miss-detection error probability while increasing the false-alarm detection error probability and vice versa. To compare the results, we have used the detection-error trade-off (DET) curves [11], with the following detection-cost function (DCF):

$$DCF = V_{miss} \times P_{miss} \times P_{true} + V_{fa} \times P_{fa} \times P_{false},$$

where $V_{miss}$ is the cost of a miss, $V_{fa}$ is the cost of a false alarm, $P_{miss}$ and $P_{fa}$ are the miss detection and false alarm probabilities, $P_{true}$ is the a priori probability of the target, and $P_{false} = 1 - P_{true}$. The optimal value, obtained from the DCF, is indicated in each plot with an "o" mark. We have used $V_{miss}= V_{fa} =1$.

The DET curves are a variant of the Receiver Operating Characteristic (ROC), that are useful when there is a tradeoff of error types (missed detections and false alarms). A verification system may fail to detect a target speaker known to the system, or it may declare such a detection when the target is not present. In contrast to the ROC curves, the DET curves present the trade-off between errors types and thus provide a better overview of the overall performance of the speaker verification system.

### 3.2 Database

For our experiments we have used the Gaudi database [7]. This database was previously used in [8] and [9]. Two subcorpora have been used:

a) ISDN: 43 speakers acquired with a PC connected to an ISDN. Thus, the speech signal is A-law encoded at fs=8 kHz, 8 bit/sample and the bandwidth is 4 kHz.

b) MIC: 49 speakers acquired with a simultaneous stereo recording with two different microphones (AKG C-420 and SONY ECM66B). The speech is in wav format at fs=16 kHz, 16 bit/sample and the bandwidth is 8 kHz.

The narrow-band signals were generated using the *potsband* routine that can be downloaded from [10]. This function meets the specifications of G.151 for any sampling frequency. The bandwidth-extension algorithm has been tuned for speech signals with 0.3-3.4 kHz bandwidth. The databases are summarized in table 1:

| Name  | BW [kHz]  | Fs | Bps | Description            |
|-------|-----------|----|-----|------------------------|
| ISDN  | [0, 4]    | 8  | 8   | Original               |
| ISDNb | [0.3, 3.4]| 8  | 8   | ISDN filtered with potsband |
| ISDNc | [0.3, 8]  | 16 | 8   | ISDNb + BW extension   |
| MIC   | [0, 8]    | 16 | 16  | Original               |
| MICb  | [0.3,3.4] | 16 | 16  | MIC filtered with potsband |
| MICc  | [0.3, 8]  | 16 | 16  | MICb + BW extension    |

Table 1: speech databases, BW = bandwidth, fs =sampling frequency (kHz), bps = bits per sample.

## 4. RESULTS

### 4.1 Results using the ISDN database

Figure 2 shows the obtained results for LPCC vectors of dimension $l$=18. We have used two different frame lengths for the bandwidth-extended signal: 30 ms (equivalent to 240 samples for a sampling rate of 8 kHz, and 480 samples for a sampling rate of 16 kHz), and 15 ms (same number of samples per frame as the original ISDN database). We can observe that the optimum is a little bit worse when the bandwidth-extension algorithm is used compared to the performance of the narrow-band system. We have obtained similar results with different vector dimensions. However, we must take into account that the ISDNc database (see table 1) has been stored in 8-bit A-law format, so quantization noise has been added during this process. For this reason, it can not be concluded that the reduction of the identification rate is due to the bandwidth-extension algorithm.

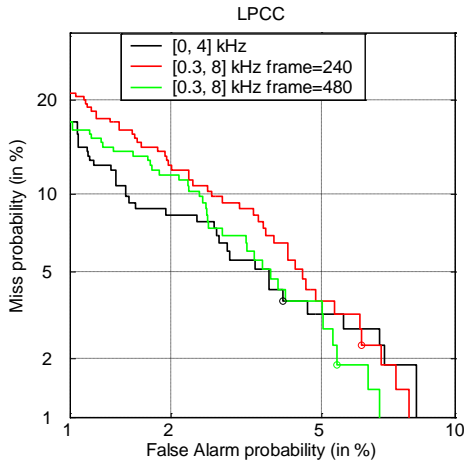

Fig. 2. DET curves for ISDN and ISDNc for LPCC-18.

Figure 3 shows the obtained results for MELCEPST parameterization for similar conditions as figure 2. We have used a suitable frame length for fft computation (256 and 512 instead of 240 and 480)

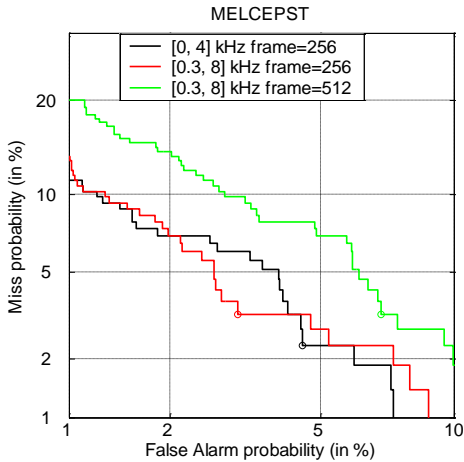

Fig. 3. DET curves for ISDN and ISDNc for MELCEPST-18.

In this case, we observe that the optimum is better than the one obtained for the ISDN database when the frame temporal length of the bandwidth extended signal is 256. We believe that this has two causes. First, the number of feature vectors is increased when the frame length is reduced, yielding a better statistical characterization of the speaker. Second, the averaging due to the mel-spaced filters reduces the effect of quantization noise, which is not the case for the LPCC parameterization.

The main conclusion of this section is that MELCEPST with a 15 ms frame length is the more suitable of the studied parameterizations. It facilitates a performance of bandwidth-extended based system similar to the narrow-band based system. Otherwise, the results are worse.

### 4.2 Results using the MIC database

Figure 4 is analogous to figure 2, but in this case the bandwidth-extended signal was extended from an unquantized narrow-band signal which better matched the training conditions of the statistical mapping used in the bandwidth extension algorithm. We observe that the bandwidth-extended signal achieves better performance than the telephone-bandwidth narrow-band signal.

The last experiment consists of the evaluation of the MELCEPST parameterizations with the different MIC databases. Figure 5 is analogous to figure 4 but using the MELCEPST parameterization. The results are similar to those obtained with the LPCC parameterization. Table 2 summarizes the numerical results for several conditions, where, as before, $l$ is the feature-vector dimension.

From table 2 we note that, using the MIC database, the MELCEPST parameterization gives better results than LPCC, especially for large $l$ values. The severe degradation in recognition performance for large dimensional LPCCs for the narrow-band database is explained by the fact that we have too few observations per parameter in the LPC estimation. Furthermore, the bandwidth-extension algorithm only improves the results achieved with the narrow-band system when the vector dimension of the LPCC is greater than $l$=18. Otherwise, the results are slightly worse. The bandwidth-extension algorithm with a MELCEPST parameterization clearly performs better than the narrow-band based speaker verification system, especially when $12 \leq l \leq 18$. Naturally, the original full-band signal (MIC database) always achieves the best performance for the typical $l$ values, because it contains more information related to the speaker.

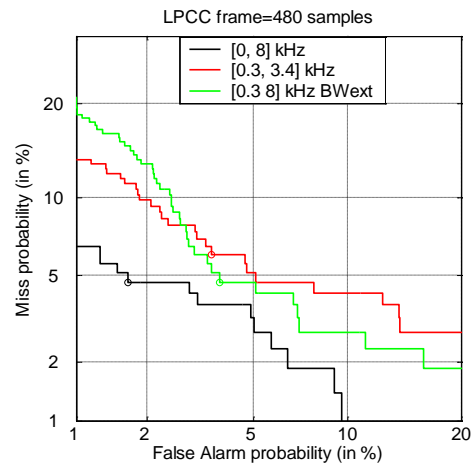

Fig. 4. DET curves for MIC, MICb, and MICc for LPCC-18.

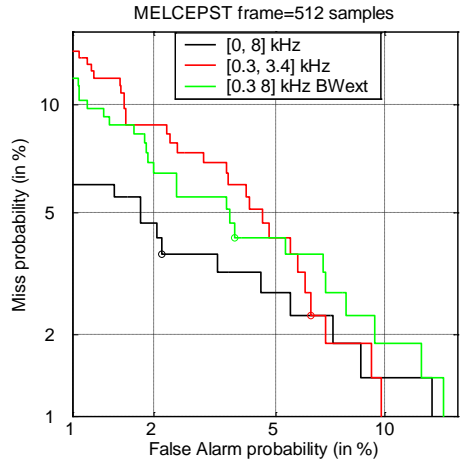

Fig. 5. DET curves for MIC, MICb, and MICc for MELCEPST-18

| $l$ | LPCC | | | MELCEPST | | |
|---|---|---|---|---|---|---|
| | MIC | MICb | MICc | MIC | MICb | MICc |
| 4 | 0.2049 | 0.1599 | 0.1912 | 0.1429 | 0.1435 | 0.1425 |
| 6 | 0.1275 | 0.1096 | 0.1006 | 0.0933 | 0.1006 | 0.0978 |
| 8 | 0.0619 | 0.0772 | 0.0797 | 0.0741 | 0.0793 | 0.0766 |
| 9 | 0.0664 | 0.0643 | 0.0643 | 0.0611 | 0.0725 | 0.0824 |
| 11 | 0.0528 | 0.0483 | 0.0596 | 0.0453 | 0.0653 | 0.0587 |
| 12 | 0.0481 | 0.0514 | 0.0527 | 0.0412 | 0.0642 | 0.0545 |
| 13 | 0.0386 | 0.0414 | 0.0447 | 0.0375 | 0.0625 | 0.0503 |
| 15 | 0.0329 | 0.0406 | 0.0471 | 0.0317 | 0.0517 | 0.0452 |
| 16 | 0.0351 | 0.0499 | 0.0428 | 0.0296 | 0.0470 | 0.0406 |
| 17 | 0.0321 | 0.0499 | 0.0493 | 0.0301 | 0.0462 | 0.0383 |
| 18 | 0.0316 | 0.0480 | 0.0422 | 0.0293 | 0.0433 | 0.0396 |
| 19 | 0.0363 | 0.0619 | 0.0458 | 0.0312 | 0.0376 | 0.0369 |
| 21 | 0.0358 | 0.0751 | 0.0481 | 0.0293 | 0.0391 | 0.0382 |
| 22 | 0.0340 | 0.0826 | 0.0470 | 0.0283 | 0.0379 | 0.0371 |
| 23 | 0.0382 | 0.0967 | 0.0495 | 0.0262 | 0.0382 | 0.0374 |
| 24 | 0.0372 | 0.1147 | 0.0478 | 0.0275 | 0.0390 | 0.0390 |
| 25 | 0.0375 | 0.1257 | 0.0503 | 0.0249 | 0.0412 | 0.0408 |
| 26 | 0.0383 | 0.1336 | 0.0523 | 0.0258 | 0.0425 | 0.0401 |
| 27 | 0.0404 | 0.1636 | 0.0512 | 0.0255 | 0.0453 | 0.0392 |

Table 2: Minimum value of the DCF for MICx databases, LPCC and MELCEPST parameterizations, and several $l$ values.

## 5. CONCLUSIONS

We have established the following conclusions:

- The studied algorithm does not introduce any damaging artifacts that affect the speaker identification rates, thus, speaker verification using bandwidth-extended signals is possible.

- The MELCEPST parameterization based speaker verification can take advantage of the bandwidth-extended speech in several situations, and outperforms the LPCC parameterization. Thus, MELCEPST is in this case the recommended parameterization for speaker verification when the system is designed for bandwidth-extended speech.

- Little improvement can be achieved with a bandwidth-extension algorithm. This is quite intuitive, because no new information is added with this system. There is just a replication of the known information. The relative improvement of the bandwidth-extended based system over the narrow-band based system is a decrease ranging between 1 and 10 percent (depending on the model order $l$) in the DCF minimum value for the MELCEPST parameterization.

## 6. References

[1] J. Leis, M. Phythian, & S. Sridharan "Speech compression with preservation of speaker identity", pp. 1711-1714, IEEE ICASSP 1997.

[2] A. Schmidt-Nielsen & D. P. Brock "Speaker recognizability testing for voice coders", pp. 1149-1152, IEEE ICASSP 1996.

[3] K. T. Assaleh "Automatic evaluation of speaker recognizability of coded speech", pp. 475-478, IEEE ICASSP 1996.

[4] M. Nilsson & W. B. Kleijn "Avoiding over-estimation in bandwidth extension of telephony speech", pp. 869-872 IEEE ICASSP 2001.

[5] J. Deller et al. "Discrete-Time Processing of Speech Signals," Prentice-Hall, 1993.

[6] F. Bimbot, L. Mathan "Text-free speaker recognition using an arithmetic-harmonic sphericity measure." pp. 169-172, Eurospeech 1993.

[7] J. Ortega et al. "Ahumada: a large speech corpus in Spanish for speaker identification and verification". pp. 773-776, IEEE ICASSP 1998.

[8] C. Alonso & M. Faúndez-Zanuy, "Speaker identification in mismatch training and testing conditions," pp. 1181-1184, IEEE ICASSP 2000.

[9] M. Faúndez-Zanuy "A combination between VQ and covariance matrices for speaker recognition". pp. 453-456, IEEE ICASSP 2001.

[10] http://www.ee.ic.ac.uk/hp/staff/dmb/voicebox/voicebox.html

[11] A. Martin, G. Doddington, T. Kamm, M. Ordowski, and M. Przybocki, "The DET curve in assessment of detection performance", pp.1895-1898, Eurospeech 1997.

[12] J. Campbell "Speaker recognition" in BIOMETRICS: Personal Identification in Networked Society. pp.165-190, Kluwer 1999.

[13] Draft New Recommendation ITU-R BS. Document 6/63-E, "system for digital sound broadcasting in the broadcasting bands below 30 MHz", 25th october 2000.